# Tailored nano-columnar La$_2$NiO$_4$ cathodes for improved electrode performance †

Alexander Stangl *[a], Adeel Riaz [a], Laetitia Rapenne [a], José Manuel Caicedo [b], Carmen Jiménez [a], Michel Mermoux [c] and Mónica Burriel *[a]

[a.] Univ. Grenoble Alpes, CNRS, Grenoble INP, LMGP, 38000 Grenoble, France.
E-mail: alexander.stangl@grenoble-inp.fr and monica.burriel@grenoble-inp.fr
[b.] Catalan Institute of Nanoscience and Nanotechnology, ICN2, CSIC and The Barcelona Institute of Science and Technology (BIST), 08193 Bellaterra, Spain
[c.] Univ. Grenoble Alpes, Univ. Savoie Mont Blanc, CNRS, Grenoble INP, LEPMI, 38000, Grenoble, France

La$_2$NiO$_4$ is a very promising cathode material for intermediate and low temperature solid oxide cell applications, due to its good electronic and ionic conductivity, together with its high oxygen exchange activity with a low activation energy. Oxygen incorporation and transport in La$_2$NiO$_4$ (L2NO4) thin films is limited by surface reactions. Hence, tailoring the morphology is expected to lead to an overall improvement of the electrode performance. We report on the growth of nano-architectured La$_2$NiO$_4$ thin film electrodes by Pulsed Injection Metal Organic Vapour Deposition (PI-MOCVD), achieving vertically gapped columns with multi-fold active surface area, leading to much faster oxygen exchange. This nano-columnar structure is rooted in a dense bottom layer serving as good electronic and ionic conduction pathway. The microstructure is tuned by modification of the growth temperature and characterised by SEM, TEM and XRD. We studied the effect of surface activity by electrical conductivity relaxation measurements in fully dense and nano-columnar La$_2$NiO$_4$ thin films of various thicknesses grown on several different single crystal substrates. Our results demonstrate that the increased surface area, in combination with the opening of different surface terminations, leads to a significant enhancment of the total exchange activity in our films with optimized nano-architectured microstructure.

## Introduction

Many aspects of modern society are strongly linked with the unrestrained availability of electrical energy in everyday life. The supply of mobile, carbon neutral energy, based on sustainable energy storage systems, however, remains a technological challenge up to date. One promising solution are electrochemical energy storage and conversion devices built up from solid-state ionic materials, such as all-solid-state batteries and reversible micro solid oxide cells (μSOC), captivating with potential high efficiencies and energy densities.

μSOC are layered heterostructures, where all SOC components (electrodes, electrolyte and current collectors) are fabricated in the form of thin films, hence significantly reducing polarization contributions from diffusion processes. Current approaches for portabilisation and commercialisation of μSOCs require low operation temperatures (≤500°C), where electrochemical performance is intrinsically reduced, with the bottleneck being the sluggish surface activity of the cathode[1,2]. Consequently significant efforts have been undertaken to improve and conserve the surface exchange activity of various electrode materials at low temperatures[3], with a strong focus on perovskite and perovskite related oxides[4], accompanied by the development and advance of new experimental techniques to access fundamental knowledge of elementary processes[5,6]. As described by Yang *et al.*, high cathode activity is a combination of i) **intrinsic activity**, defined by physicochemical properties including charge carrier densities and their mobilities, defect formation enthalpies, *etc.* and ii) **apparent activity**, *e.g.* the total number of active sites[7]. The former is an intensive quantity, while the latter is of extensive nature, leading to two different optimisation tasks. Strategies to enhance intrinsic activity include the investigation of new material structures[8–10] and compositions[11,12], engineering interfaces[13], strain states[14–16] and defect landscapes[17–19] and the decoration of surfaces with catalytic nanoparticles via infiltration[20,21], heterostructuring[22] and exsolution[23,24] to alter surface termination and surface electronic states. On the other hand, apparent activity can be tuned by extending the active electrode area due to enhanced ionic conductivity (mainly in μm thick porous electrodes), or via increasing the specific surface area (SSA, *e.g.* total topographic surface area/in-plane area) by three dimensional, nano-architectured morphologies in thin film electrodes, such as nanofibrous[25,26], mesoporous[27,28] and columnar structures[29]. However, a distinct separation between intrinsic and apparent activity is not always possible, as in nano-engineered (composite) materials these two effects are strongly coupled.

Tailoring the electrode topography requires insight into the growth mechanisms during film deposition, as has been studied since decades[30–32]. The initial nucleation stage during heterogeneous growth occurs either as two dimensional layer-by-layer (Frank-van der Merwe) or three dimensional island (Volmer-Weber) or layer-plus-island growth (Stranski-Krastanov) and depends on the interface energy between substrate atom and film adatom and supersaturation[33]. Chemical vapour deposition techniques are generally thought to possess high supersaturation leading to Volmer-Weber growth, linked to a higher defect density and surface roughness. The nucleation phase is followed by coalescence of the nuclei until the substrate surface is buried. The subsequent film growth is governed by the interplay of mainly four different processes, namely: shadowing, surface and bulk diffusion and recrystallization[32]. The temperature dependent prevalence of one of these four mechanisms causes the formation of a distinct microstructure, allowing to construct an *ideal* structure zone model[30], by applying the Thornton model to a chemical vapour deposition process:

**Zone I** at low temperature is characterised by low mobility of adatoms within the surface and the bulk leading to a porous structure with finely sized columns, directly evolving from the nucleation sites. Shadowing has significant influence on the growth.

**Zone T**, in the next temperature interval, is determined by the onset of surface diffusion. The competitive growth of different grains causes the emblematic V-shaped columnar texture of grains, with a homogenous small grain size at the substrate interface, increasing with film thickness. Porosity can persist as long as bulk diffusion is limited.

In **Zone II** bulk diffusion is activated by sufficiently high temperatures giving rise to grain coarsening and a homogenous columnar structure across the film thickness, whereat the grain size increases with temperature.

At even higher temperatures, recrystallization tends to minimise surface and interface energies, resulting in a restructuring of the crystal structure (**zone III**).

It is noted, that a real structure zone model will be significantly affected by any kind of impurities within the growing layer, supersaturation, oxygen concentration, *etc*[31].

A columnar structure with extended surface area can be achieved through growth in zone 1 and zone T, as experimentally found for $La_xSr_{1-x}Co_yFe_{1-y}O_{3-\delta}$ (LSCF) based cathodes deposited by PLD[29,34,35] and magneton sputtering[25], where the specific surface area was multifold, *e.g.* from 1 for dense to 26 for porous LSCF films. This increase in apparent activity resulted in an inverse proportional drop of the area specific resistance[36], proving the potential of this nano-architectural approach.

As material of interest in this study, we have selected $La_2NiO_4$ (L2NO4), a member of the Ruddlesden-Popper family with $K_2NiF_4$ structure. This is one of the most promising material classes currently under investigation for SOC oxygen electrodes and oxygen permeation membranes, due to fast ionic diffusivity, good electrical conductivity and salient catalytic activity towards the oxygen reduction reaction (ORR), with low activation energy and wide chemically stability. The thermal expansion coefficient matches with commonly used electrolyte materials, a prerequisite for mechanical stability. $La_2NiO_4$ has been widely studied in the form of thin, dense films and the rate determining step has been generally ascribed to a surface mechanism[37,38]. Ionic transport properties in $La_2NiO_4$ and other rare earth nickelates ($Pr_2NiO_4$ and $Nd_2NiO_4$) are highly anisotropic, with oxygen diffusion being 100 to 1000× faster within the *ab*-plane[39–42] Likewise, surface oxygen kinetics were found to be generally faster in the *ab*-plane as compared to the *c*-axis[39–42], although a certain history-dependence has also been reported[41]. While theoretical calculations show that the Ni terminated (001) surface is supposedly the most active one[43,44], followed by the (111), (100) and the La terminated (001) surface, it was experimentally observed that Ni is highly deficient in the outermost surface layers[45,46]. A recent study on epitaxial LSCF thin films with different orientations revealed 3× faster kinetics of the (111) surface as compared to the (001) one, due to a higher number of active sites, while being dominated by the same reaction mechanism[47].

In agreement with theoretical calculations and experimental observations in many perovskite systems, tensile volumetric strain in (100) oriented epitaxial L2NO4 thin films was shown to significantly impact surface exchange kinetics, resulting in faster exchange coefficients, $k^q$, for thinner, higher strained films[48]. However, it should be noted that earlier studies by Isotope Exchange Depth Profile (IEDP) did not find any dependence of the tracer surface exchange with thickness in epitaxial (001) oriented L2NO4[40].

Pulsed Injection Metal Organic Vapour Deposition (PI-MOCVD) is a highly interesting and versatile deposition technique suitable for industrial application, as it is able produce well controlled thin films of complex compositions with good quality and precision, and covering large surfaces areas. In this work we report for the first time on an innovative strategy to tune intrinsic and apparent surface activity of $La_2NiO_4$ thin films deposited by PI-MOCVD, by tailoring the nanostructure. This enhancement is based on the one hand on modifications of the morphology, *e.g.* due a transformation from a dense layer to a nano-columnar structure, controlled via the deposition temperature and film thickness. The microstructure consists of nano-columns rooted in a dense layer at the interface with the substrate, which serves as conduction and diffusion pathway and guarantees proper current collection for SOC application. This architecture leads to a significant increase in surface to volume ratio and, consequently for a surface limited reaction, higher apparent exchange activity. On the other hand, the exposure of new surface terminations at the lateral side of the nano-columns and the loss of crystal texture with film thickness, results in enhanced intrinsic activity of the cathode material. The enhancement of the oxygen exchange kinetics is studied by electrical conductivity relaxation measurements, showing the potential of this nano-architecture design strategy to optimise the microstructure for high exchange activity.

## Experimental

### Thin film preparation

L2NO4 thin films are grown by PI-MOCVD on 10×10mm $LaAlO_3$ (LAO), $SrTiO_3$ (STO) and YSZ single crystal substrates with (100) orientation. La and Ni commercial TMHD precursors are dissolved in m-xylene with a La/Ni ratio of 5 and a total concentration of 0.02mol/l. The depositions are carried out in an $Ar/O_2$ (34/66%) atmosphere with a total pressure of 5 Torr, while deposition temperature is varied between 650 and 750 °C. Film thickness is controlled by the number of injected pulses and was varied between 33 and 540 nm.

### Sample characterisation

Phase and orientation of derived thin films are characterised by X-ray diffraction (XRD) in $\theta - 2\theta$ configuration using a Bruker D8 Advance series II diffractometer ($CuK_\alpha$ radiation). The surface homogeneity was studied using a FEG Gemini SEM 500 microscope in secondary (inlens) and backscattered electron mode with an accelerating voltage of 3 kV in high vacuum. Post image analysis was supported by the deep learning feature of MIPAR software. The surface morphology was further investigated by atomic force microscopy (AFM) in tapping mode using a Bruker Icon high performance AFM. Film thickness was obtained via X-ray reflectometry and transmission electron microscopy (TEM) analysis.

L2NO4/LAO specimens were prepared in cross-section by the semi-automated polishing tripod technique with the MultiPrep™ system (Allied High Tech Products, Inc.). The PIPS II from GATAN was used for the final polishing. TEM and high resolution TEM (HRTEM) images were recorded with a JEOL JEM 2010 LaB$_6$ microscope operating at 200 kV with a 0.19 nm point-to-point resolution. The local structural properties of L2NO4/LAO films were further investigated by using an automated crystal phase and orientation mapping (ACOM) with a precession system (ASTAR) implemented in a JEOL 2100F FEG microscope. The crystal phase and orientation maps were obtained by precession of the primary electron beam around the microscope's optical axis at an angle of 1.2° while collecting the electron diffraction patterns at a rate of 100 frames/s with a step size of 2 nm. In this technique, the incident electron beam was a few nanometers in size and was precessed to reduce the dynamical effects and to enhance the indexing quality. The electron beam was simultaneously scanned over the area of interest to record an electron diffraction pattern at each location. For each point the local experimental electron diffraction patterns were compared with the complete set of theoretical diffraction patterns, which were computed for every expected crystalline phase and for a large number of orientations. The best match between the experimental and theoretical electron diffraction patterns permitted identification of both crystalline phase and orientation with a high precision. The crystal structures used for identification included La$_2$NiO$_4$, LaNiO$_3$, La$_2$O$_3$ (cubic and hexagonal) and LaAlO$_3$. Reliability is expressed by the lightness value in the index map, *e.g.* bright corresponds to high certainty.

**Functional analysis**

Functional properties were studied by electrical conductivity and electrical conductivity relaxation (ECR) measurements. Samples were cut with a water cooled diamond saw into pieces of 5×5 mm. Electrical contacts were fabricated in the corners using Ag paint. For *in situ* characterisation, the samples were placed onto a 1/2" ceramic heating stage in a high temperature cell (Nextron). The surface temperature of the heater was calibrated beforehand using a Pt100 thermocouple and is assumed to be equivalent to the thin film temperature. The cell is equipped with electrical probes which are mechanically pinned onto the Ag electrodes and electrical conductivity is measured in Van-der-Pauw configuration using a small excitation current of 100 µA and 1 µA for L2NO4/(LAO or STO) and L2NO4/YSZ, respectively. The sample is annealed in a dynamic, dry, mixed O$_2$/N$_2$ atmosphere at 1 atm and constant flow rate of 1000 ml/min, whereas the oxygen partial pressure is controlled via the flow ratio of highly pure oxygen to nitrogen. The *p*O$_2$ is measured at the exit of the chamber using a Rapidox 2100 gas analyser. Temperature, *p*O$_2$ and conductivity are automatically controlled and continuously recorded using Labview. The evolution of the conductivity is followed during isothermal annealings as a function of consecutive steps in *p*O$_2$ between 210 and 10 mbar. The commonly used solution to Fick's second law of diffusion for a thin film with a linear surface exchange, is applied and proportionality between the transients of conductivity, $\sigma$ and oxygen concentration, $c$, is assumed. One obtains[49]:

$$\frac{\sigma(t) - \sigma_\infty}{\sigma_0 - \sigma_\infty} \propto \frac{c(t) - c_\infty}{c_0 - c_\infty} = \sum_{i=1}^{\infty} A_i e^{-t/\tau_i} \qquad \text{Equation 1}$$

where indices indicate initial ($t = 0$) and equilibrium ($t = \infty$) state and

$$A_i = \frac{2\delta^2}{\beta_i^2(\beta_i^2 + \delta^2 + \delta)} \qquad \text{Equation 2}$$

$$\delta = \beta_i \tan \beta_i = \frac{dk_{\text{chem}}}{D_{\text{chem}}} \qquad \text{Equation 3}$$

$$\tau_i = \frac{d^2}{D\beta_i^2} \qquad \text{Equation 4}$$

with $d$ being the film thickness, $A_i$ the weight for the corresponding time constant, $\tau_i$ and $\beta_i$ the $i$'th positive root of the transcendental Equation 3. For surface controlled processes, all time constants for $i \geq 2$ vanish, resulting in a single saturation time

$$\tau = V/A/k_{\text{chem}}^\delta \qquad \text{Equation 5}$$

with $A$ being the active, exposed surface area and $V$ the sample volume. For the case of dense flat films, this corresponds to $\tau = d/k_{\text{chem}}^\delta$.

In the mixed and diffusion limited regime, $A_i \neq 0$ for all $i$. However, the value of $A_i$ strongly decreases with increasing index, allowing one to focus on the first two time constants, having the highest values. For $t \to \infty$ the normalised oxygen concentration of the solely diffusion limited process can be approximated by a single exponential decay[50] with $\tau_1 = 4d^2/(\pi^2 D_{chem})$ and $A_1 = 8/\pi^2$.

**Finite element method (FEM) simulations**

FEM simulations are performed to analyse the thickness dependence of the exchange time constants for different sample geometries using COMSOL 5.5 on a 3D model of a single nano-column. A schematic is shown in Figure 4. Oxygen exchange takes place at the top and lateral surfaces of the nano-rods, with $k^\delta$ as obtained for the dense L2NO4/YSZ film. The diffusion coefficient is assumed constant but anisotropic with $D_{\text{chem,ab}} = 100 D_{\text{chem,c}}$[40]. Time constants of the exchange process are obtained via fitting of the average oxygen concentration of the dense bottom layer as a function of time using one (surface limited regime) or two (diffusion limited process) exponential decays.

**Results**

**Microstructure**

Orientation and microstructure of L2NO4 thin films for different thickness are analysed by XRD using ICDD reference 04-015-2147, as shown in in Figure 1 for L2NO4/LAO (red curves). XRD patterns for all studied films (4 different thicknesses for each of the 3 studied substrates) can be found in ESI-Figure 1. Samples deposited on LAO and STO, both show preferential growth along the *c*-axis (strong (00l) reflections) due to the small mismatch between the tetragonal L2NO4 in-plane cell parameter (3.868 Å) and the cubic unit cell of the substrate (3.791 Å and 3.905 Å, respectively), with minor contributions from the L2NO4 (103) and (110) orientations. However, with increasing thickness a contribution along the (200) direction arises, as exemplarily represented for L2NO4/LAO. For the thickest films, comparing the XRD peak areas of the (200) and (00l) reflections, the overall *ab*-fraction within this film is estimated to be about 20 %. This is caused by a loss of epitaxial growth with increasing film thickness, as was previously reported for L2NO4/STO thin films deposited by PLD[51]. In some samples, a small excess of La is observed via the presence of the $La_2O_3$ phase at $2\theta=30$ °. Polycrystalline growth with preferential orientation along the (110), (100) and (103) direction is observed for films deposited on YSZ with cell parameter of 5.144 Å, also shown in Figure 1 (green diffraction pattern). The two epitaxial and the preferentially-oriented polycrystalline systems allow us to study the influence of orientation on exchange kinetics.

Epitaxial growth on LAO or STO introduces tensile or compressive strain, respectively, in very thin layers. However, the *c*-axis lattice parameter of the as-deposited L2NO4 films, calculated via the Nelson – Riley formula, does not reveal a substrate dependent thickness dependence (between 33 and 540 nm), see ESI-Figure 2. Therefore, we assume that any substrate induced strain is released by the formation of extended structural defects, already for the thinnest films studied.

The L2NO4/LAO and L2NO4/STO samples deposited at 650 °C, consist of homogenous, square and rectangular shaped grains as observed by SEM analysis, Figure 2. With increasing thickness the structure evolves from a dense layer to a columnar one, with the individual grains becoming vertically separated. This is expected to lead to an increase of the specific surface area and the surface to volume ratio (SVR). Additionally, the apparent average grain size increases, while the grain size distributions becomes broader (ESI-Figure 3). In agreement with the structure zone model, the formation of the columnar structure can be prevented by increasing the deposition temperature, *e.g.* a 200 nm thick film deposited at 750 °C possesses a dense surface, see ESI-Figure 4. L2NO4 thin films grown on YSZ single crystal substrates show the same tendency of increasing grain size and evolution towards separated columns with thickness, but the grain shape does not show the same rectangular features. The surface morphology from AFM imaging shows increasing roughness with increasing film thickness, ESI-Figure 5.

Cross-section TEM analysis of the 33 nm thick film is shown in the top panel of Figure 3(a-c). The structure consists of closed packed columnar grains and a smooth, flat surface. The 200 and 540 nm thick films, shown in Figure 3(d-f) and Figure 3(g-i), respectively, display a different microstructure featuring two distinct zones, namely a dense interfacial layer on top of the substrate, followed by a nano-columnar structure. The thickness of the dense bottom layer, $d_\mathrm{d}$, shows some local variation with an average thickness of approximately 80 and 100 nm, respectively. Such a dense layer was described previously and seems to be a common feature for the growth of oxides by physical and chemical vapour deposition techniques[29]. In μSOCs it is expected that this dense layer serves as out-of-plane oxygen diffusion and in-plane electronic (holes) conduction pathway, with good adherence to the electrolyte. The grains in the columnar top region are vertically gapped by a few nanometres and cavities or nano-pores are present in the material, in which the gas is expected to enter (red arrows in Figure 3(e&h)). For the 540 nm thick sample, V-shaped grains can be found, as exemplarily marked for one of the grains in Figure 3(g), which is characteristic for zone T type growth. Additionally for the thickest film, a high density of kinks and edges is found on the lateral surface of the nano-columns, marked with blue arrows in Figure 3(i). These regions are potential hosts to highly active sites, as surface electronic states are thought to be significantly varied in proximity to these defects. STEM EDX images are displayed in ESI-Figure 6, exemplarily for the 540 nm thick L2NO4/LAO sample, showing a homogeneous La and Ni distribution throughout the film.

Based on our TEM observations, we designed a simplified geometric model shown in Figure 4, similar to the one developed initially by Plonczak *et al*.[29]. This model was used for the COMSOL simulations and for the estimation of the thickness dependence of the specific surface area. The parameters used to model each of the films, such as thickness, grain size and dense layer thickness are obtained by SEM and TEM observations and are summarised in Table 1.

Automated crystal phase and orientation mapping in TEM (ASTAR) has been used to further study the film microstructure on a nanoscale, shown in Figure 5. As already shown by XRD, $La_2NiO_4$ is the dominant phase across the whole film thickness. Only in the upper half, a small quantity of impurities in the form of $La_2O_3$ inclusions could potentially be present (low reliability index). Ambiguity, and hence low reliability, in near surface regions is likely caused be the superposition of several small grains within the thickness of the TEM lamella. The orientation map shows a thin *c*-axis oriented interfacial layer on top of the substrate. This epitaxial layer is in some regions limited to about 10 nm, followed by (100) oriented L2NO4, which in the observed region makes up more than 50 % of the thin film volume.

**Functional properties**

**Electrical conduction.** High electrical conductance and good current percolation are key for excellent electrode performance by enabling proper current collection and hence maximising the active surface area. At 375 °C the conductivity of the dense, 33 nm thick film is about 40 S/cm. This is a factor 2 to 10 lower than reference values for L2NO4 thin films from literature[52,53], likely caused by impurities due to excess La in the structure.

The high temperature in-plane conductance (Van-der-Pauw) increases with increasing film thickness, as shown for the different samples in Figure 6(a). While in the dense, 33 nm thick film the full sample volume contributes to the observed conductance, in the thicker, nano-columnar structured layers only the dense, bottom part of the cross section is active in the current percolation. Hence, for $d$>33 nm, the measured conductance falls below the extrapolated conductance from the thinnest sample. As a first approximation, assuming constant conductivity throughout the whole sample thicknesses (neglecting effects from oxygen stoichiometry and other kind of defects), we can estimate an average thickness of the dense, conduction pathway, as shown in Figure 6(b). Values are in reasonable agreement with TEM observations, where some local differences were found.

**Oxygen activity.** To gain insight into oxygen exchange activity, we performed electrical conductivity relaxation (ECR) measurements at various temperatures on L2NO4 thin films of different thicknesses on the three substrates: LAO, STO and YSZ. ECR is a standard technique used to analyse the oxygen exchange activity of oxide materials, and has been extensively used to study the oxygen kinetics of L2NO4 thin films[37,54] and bulk samples[55,56]. The process of oxygen incorporation at the surface of a solid material can be expressed by the oxygen reduction reaction (ORR):

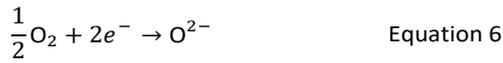

$$\frac{1}{2}O_2 + 2e^- \rightarrow O^{2-} \qquad \text{Equation 6}$$

which for the case of L2NO4, in which oxygen is incorporated in interstitial sites, can also be written following Kröger–Vink notation, as:

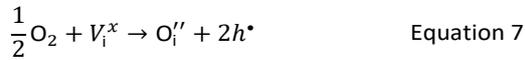

$$\frac{1}{2}O_2 + V_i^x \rightarrow O_i'' + 2h^\bullet \qquad \text{Equation 7}$$

being $O_i''$ and $V_i^x$ the oxygen interstitial and its vacancy, respectively, and $h^\bullet$ an electron hole in the valence band. Factors governing the ORR are i) temperature, ii) the surface exchange coefficient, $k_{chem}^\delta$ and iii) the chemical diffusion coefficient, $D_{chem}$, iv) the (electro-) chemical driving force, e.g. the oxygen chemical potential difference, $\Delta\mu_O$, between initial state and final equilibrium, v) the surface area, where the reaction takes place, as well as vi) the sample volume, which defines the amount of oxygen ions which need to pass through the surface to compensate $\Delta\mu_O$.

Representative normalised conductivity transients of L2NO4 thin films at 375°C after a change of $p$O$_2$ from 10-250 mbar are shown in ESI-Figure 7. The oxidation time constants are shown in an Arrhenius plot in Figure 7. For all samples thermal activation is observed, with shorter time constants at higher temperatures.

We stress the fact that saturation times drastically decrease with increasing film thickness, as depicted for one temperature (T=375 °C) in Figure 8. Independent of the substrate, saturation times are one order of magnitude shorter for the thickest films as compared to the dense 33 nm thick ones. A similar, but much weaker dependence of saturation times on thickness was observed previously in L2NO4 thin films and was ascribed to increased roughness in thicker films[54].

When comparing films grown on LAO and STO with YSZ, one can evaluate the influence of orientation on the exchange activity. While films on LAO and STO exhibit similar saturation times, samples grown on YSZ have shorter time constants. The thickness and substrate dependence will be discussed in the following.

## Discussion

To get a better understanding of the influence of the different microstructures on the oxygen exchange kinetics, COMSOL simulations have been performed using the model shown in Figure 4 for four different scenarios, with the geometric model parameters given in Table 1. The resulting thickness dependences of saturation times, corresponding to values obtained from ECR measurements, are shown in Figure 9 and compared to the ones found experimentally for L2NO4/YSZ.

**1) Diffusion limited, dense film:** if the Biot number is sufficiently large (Bi = $t \cdot k_{chem}/D_{chem}$ > 30), the exchange process is diffusion limited[49] and saturation times are approximately proportional to the square of the thickness.

**2) Diffusion limited, nano-columnar structure:** also for this case FEM simulations reveal an increase of saturation times with thickness, however, a much weaker dependence is obtained as compared to the dense, diffusion limited scenario.

**3) Surface limited, dense layer:** this is the only case with an exact analytic solution, where the saturation time increases linearly with film thickness, $\tau = d/(k_{chem}^\delta)$.

**4) Surface limited, nano-columnar structure:** this scenario strongly depends on the geometric model. In the realm of surface limitation, $\tau$ is inverse proportional to the surface to volume ratio: $\tau = (V/A)/(k_{chem}^\delta)$. Hence, the introduction of surface roughness or open porosity can have drastic impact on measured saturation times, as can be clearly seen for case 4 in Figure 9, being the only situation where the time constant is expected to decrease with thickness, as experimentally found for the studied thin films (see also Figure 8). In the presence of nano-columns, the active surface area is strongly enlarged, consequently increasing the total oxygen ion flux through the surface. This shows the potential of nano-tailoring surfaces and morphologies to enhance the apparent oxygen exchange activity of electrode materials. In the present model, grain size and its evolution with thickness firmly affect the surface to volume ratio, while the width of the gap between two grains has a minor influence. Simulations were performed using the grain size distribution obtained by SEM analysis. However, the enlarged surface alone cannot fully account for the strong decrease of the time constant with thickness found experimentally, which is discussed in the following.

The influence of nano-columnar growth on the surface exchange coefficient, $k_{ex}$, calculated using Equation 5 and the SVRs given in Table 1, is depicted in Figure 10. Exchange coefficients for the nano-columnar films are significantly faster than the ones of the dense sample. We suggest that the enhancement of intrinsic activity is due to the exposure of additional crystal orientations and associated changes in surface termination and surface chemistry, and possibly highly active nano-features such as kinks and edges being numerously present within the lateral surface. The gain of activity related to the enlarged surface area, can be depicted using the apparent exchange coefficient, equivalent to the surface exchange coefficient required for a dense layer (of same thickness) to exchange oxygen as fast as a nano-columnar one.

With this in mind, one can return to the differences found in Figure 7(b), between films of different orientation. While films on LAO and STO exhibit preferential *c*-axis growth (001), particularly for the thinnest films, L2NO4 layers on YSZ grow mainly along the (110) and (100) direction. From previous experimental and theoretical results it is known that the (100) terminated surface has higher intrinsic activity[40,43] as compared to the (001) orientation. This can explain the shorter time constants obtained for L2NO4/YSZ with respect to films on LAO and STO. Taking into account that with increasing film thickness, also films on LAO and STO show an increasing presence of the (100) orientation, it is expected that this contributes to the faster oxygen exchange in these films.

Another potentially influencing factor, currently being investigated within our group, is the amount of Ni present directly in the top surface layers. A-site (La, Sr) cation domination was found in several perovskite-based, dense oxides[57], and in particular in $(La,Sr)_2NiO_4$ single crystals with cleaved surfaces[45] and $La_2NiO_4$ bulk[57]. However, the La and Ni cation distribution could be strongly affected by the nano-columnar structure with its small grain size, with implications on the surface chemistry of the material, which is yet to be confirmed.

## Conclusions

In this work we have demonstrated how to tune apparent and intrinsic activity of $La_2NiO_4$ thin films by tailoring its nano-structure. Samples were synthesized by PI-MOCVD on different single crystal substrates with thickness ranging between 33 and 540 nm. We were able to tune the morphology of the films by varying the deposition temperature (650 - 750 °C) in agreement with the structure zone model, to achieve a nano-columnar microstructure with open porosity and significantly enlarged specific surface area at lower deposition temperatures. ECR measurements revealed remarkably enhanced surface activity in thicker, nano-architectured films, as compared to thin, dense ones. While the larger surface area plays an important role, it does not account for the total activity increase. Hence, apparent as well as intrinsic activity was improved. We concluded that the latter was enhanced by changes in the grain orientation and the exposed surfaces with distinct, faster exchange coefficients, while additional features on the lateral side of the nano-columns such as kinks and edges are thought to also improve oxygen exchange, as these defects are expected to strongly modify surface electronic states involved in the reaction mechanism. Overall, the nano-columnar growth of L2NO4 by PI-MOCVD resulted in a substantial enhancement of the oxygen exchange activity and opens up a new route towards the optimisation of electrode materials for intermediate to low temperature devices.

## Author Contributions

Conceptualization: MB, AS; Investigation: AR, AS, LR, JMC; Formal Analysis: AR, AS, LR; Methodology: AS, AR, CJ, MB; Writing – original draft: AS; Writing – review & editing: AS, MB, AR, CJ, MM.

## Data availability

All sample datasets and materials related to this work is made available under CC BY 4.0 license in the zenodo repository: 10.5281/zenodo.5564457

## Conflicts of interest

There are no conflicts to declare.

## Acknowledgements

This work has received funding from the European Union's Horizon 2020 research and innovation program under grant agreement no. 824072 (Harvestore). The authors further acknowledge financial, scientific and technical support from CNRS, Grenoble INP, UGA and the scientific and technical assistance of Gilles Renou for ASTAR technique of the CMTC characterization platform of Grenoble INP, which is supported by the Centre of Excellence of Multifunctional Architectured Materials (LabEx CEMAM).

# ARTICLE

| Total film thickness | Grain size | Thickness dense layer | SVR |
|---|---|---|---|
| nm | nm | nm | µm$^{-1}$ |
| 33 | 28±8 | 33 | 30 |
| 100 | 46±12 | 60* | 44 |
| 200 | 50±18 | 80±20 | 52 |
| 540 | 55±27 | 100±20 | 61 |

Table 1: Thin film parameters obtained by SEM and TEM analysis for dense (33 nm) and columnar structured thicker films on LAO and the resulting surface to volume ratio. *estimated via electrical measurements (see main text)

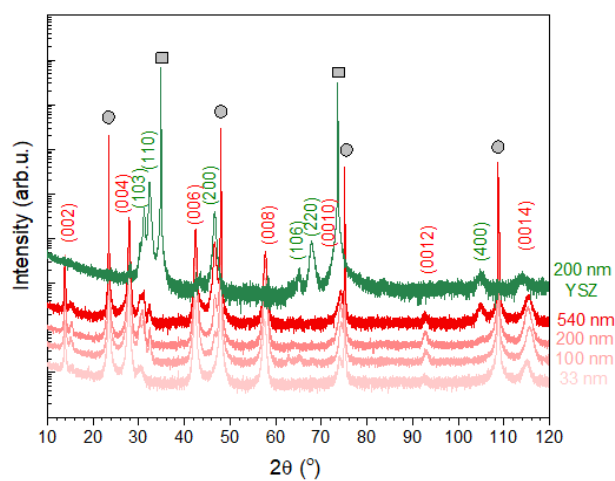

Figure 1: XRD pattern of L2NO4 thin films of various thicknesses deposited at 650 °C on LAO (red curves). Additionally, the 200 nm L2NO4/YSZ sample is shown (green), while XRD data for all thicknesses on all substrates are compared in ESI-Figure 1. Substrate peaks are marked by grey circles (LAO) and squares (YSZ). The observed L2NO4 planes are labelled according to ICDD reference 04-015-2147. Films on LAO possess a preferential growth direction along the (001) direction. With increasing thickness, an additional orientation along the (100) direction is detected. For L2NO4/YSZ, polycrystalline growth preferentially-oriented along the (110), (100) and (103) direction is observed.

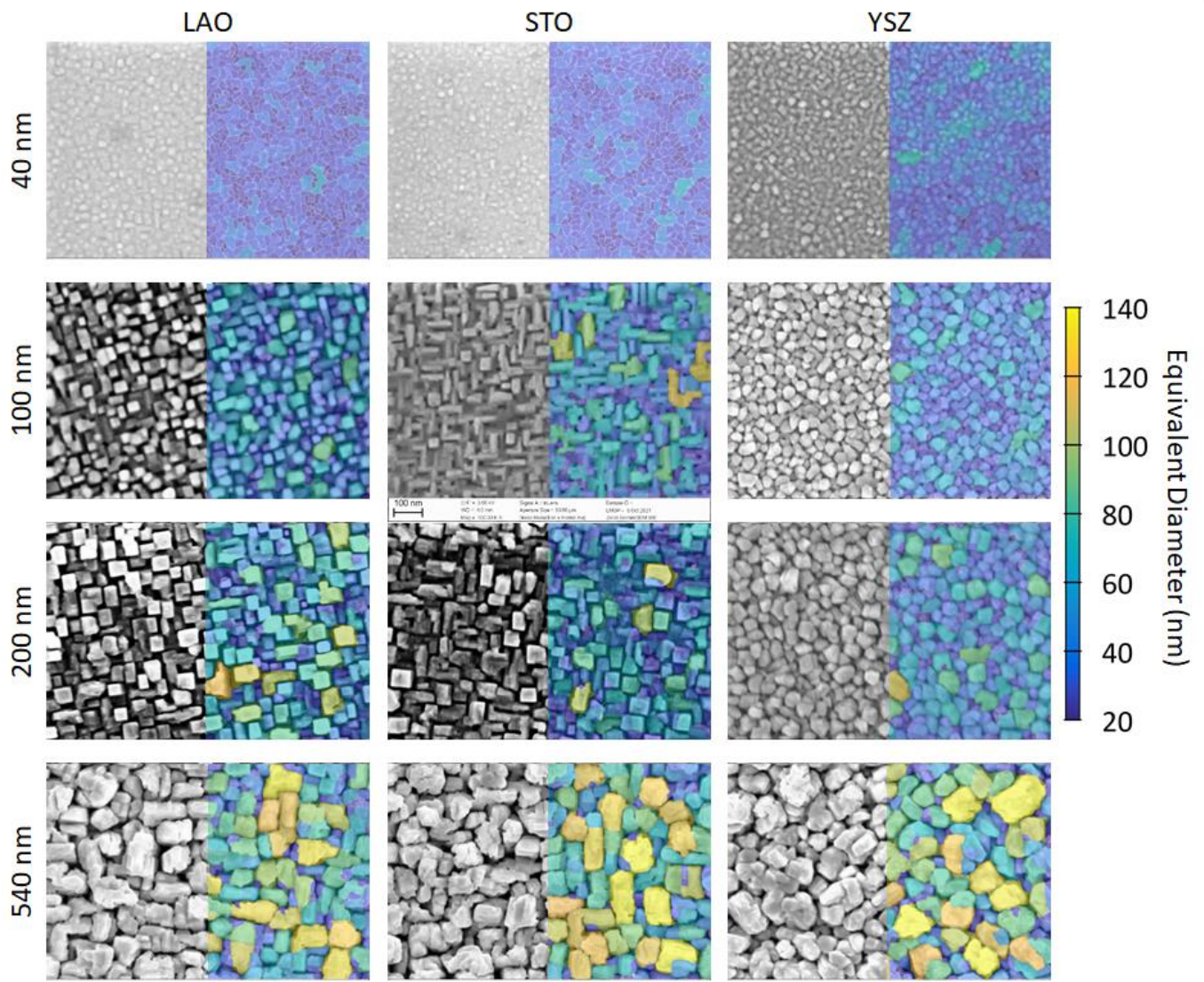

Figure 2: Surface morphology of L2NO4 thin films: SEM analysis of L2NO4 films of different thicknesses deposited on top of LAO, STO and YSZ single crystals at 650 °C. False colour indicates grain size using the equivalent diameter of a circular grain of the same area. Corresponding histograms are shown in ESI-Figure 3.

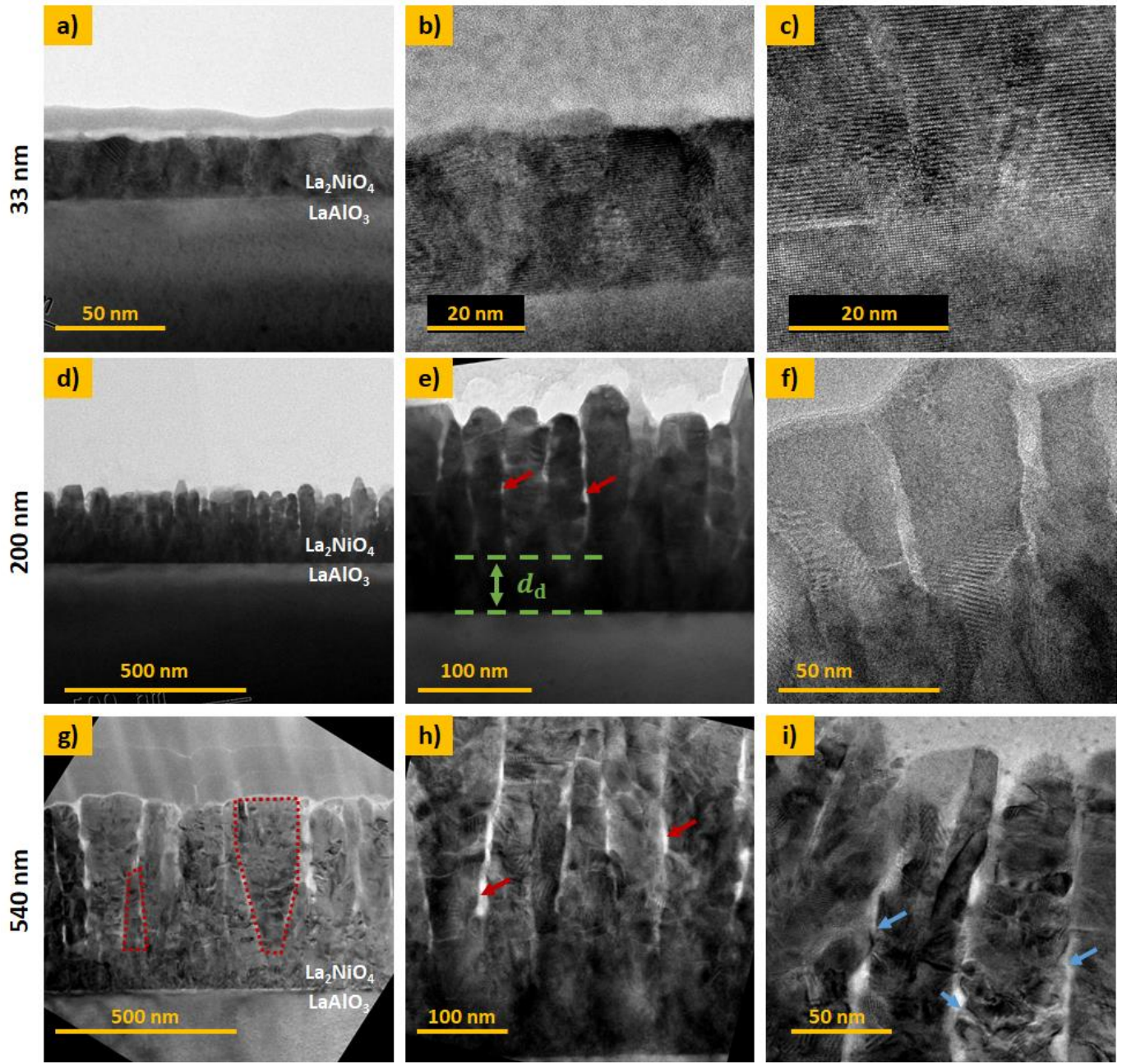

Figure 3: Cross-section TEM of dense 33 nm thick L2NO4 (a-c) and nano-columnar structured 200 nm (d-f) and 540 nm (g-i) thick L2NO4 on top of LAO. A dense interfacial layer of thickness $d_d$ is observed below the nano-columnar growth, as indicated in e). Red arrows in e) and h) mark porosity between vertically gapped nano-columns. A high number of kinks and edges is found at the lateral surface of the nano-columns in the 540 nm thick film, which are potential sites for high activity due to local variations in surface electronic states, see blue arrows in i).

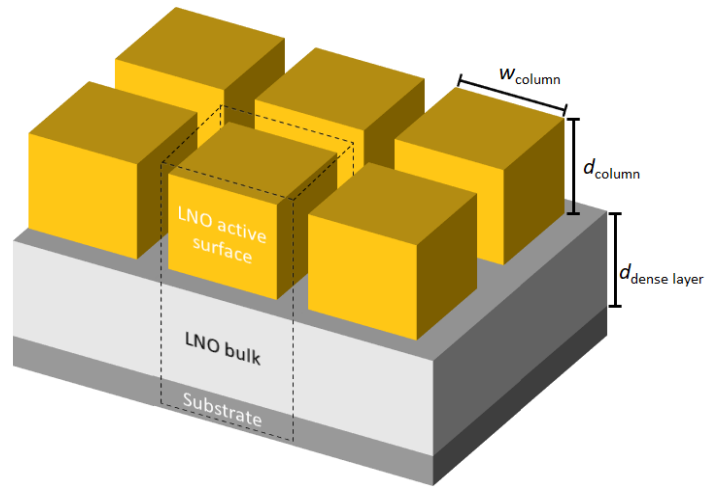

Figure 4: Simple geometric model for finite element (COMSOL) simulations to study influence of geometrical parameters on oxygen exchange activity.

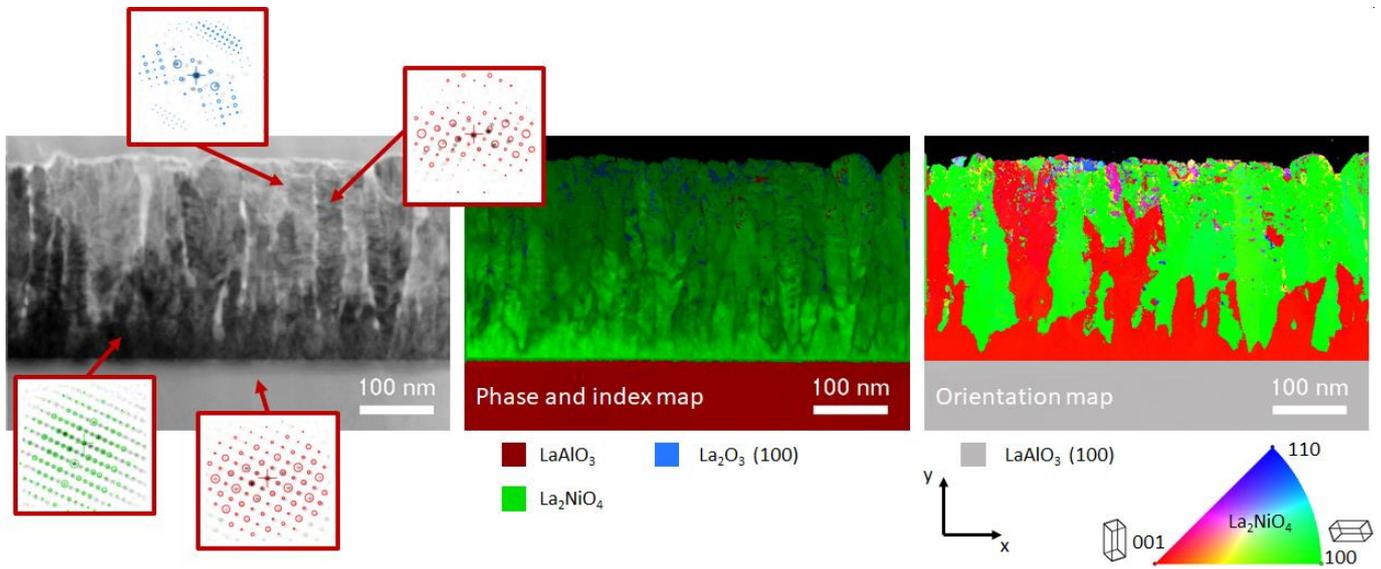

Figure 5: ASTAR TEM analysis of 540 nm L2NO4/LAO sample. a) Bright field TEM image. The insets show local electron diffraction patterns, open circle correspond to the reference diffraction patterns of the best phase match. b) ASTAR map representing the crystalline phases, lightness of the colour corresponds to reliability of the identified phase. c) ASTAR orientation map, colour code corresponds to planes perpendicular to the $y$-axis.

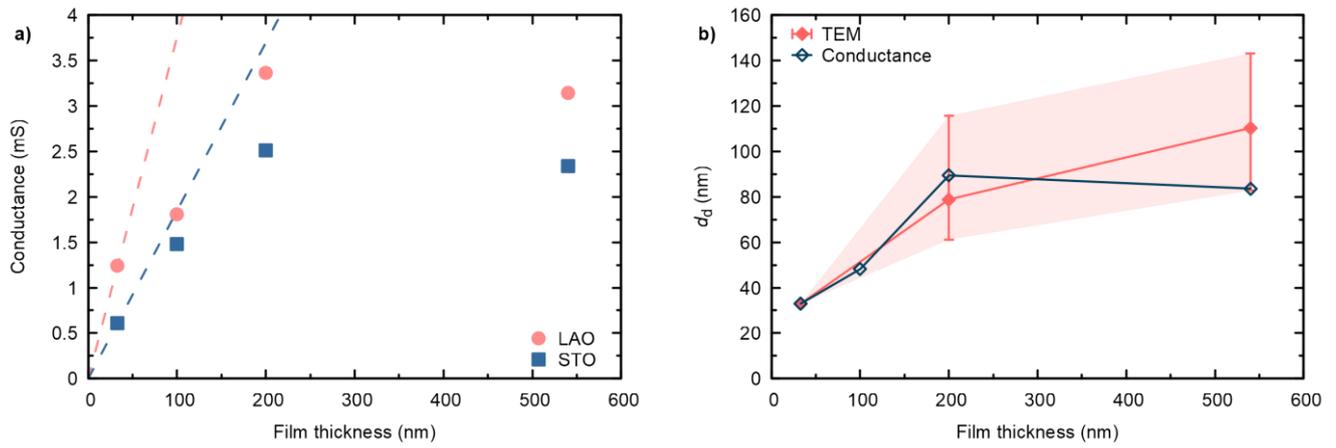

Figure 6: (a) Measured equilibrium conductance by Van-der-Pauw measurements at 375 °C in high $pO_2$. Dashed lines are extrapolations of the expected conductance of a fully dense layer. (b) Thickness of dense bottom layer, $d_d$, for L2NO4/LAO samples, estimated from conductance measurements assuming constant conductivity and via TEM measurements. Filled area indicates observed variations of the dense layer thickness by TEM, e.g. minimum and maximum values.

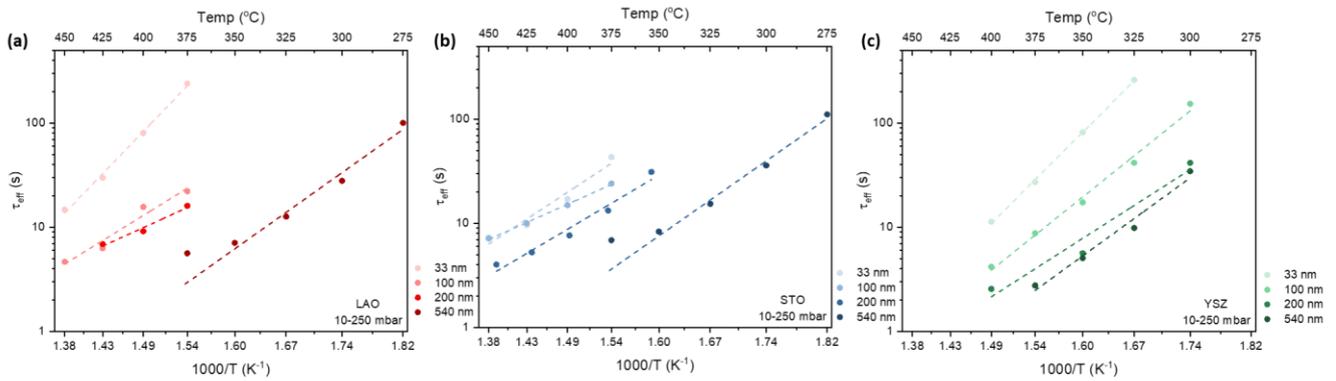

Figure 7: Arrhenius plot of oxidation saturation times from ECR measurements for L2NO4 films with different thickness on a) LAO, b) STO and c) YSZ single crystal substrate.

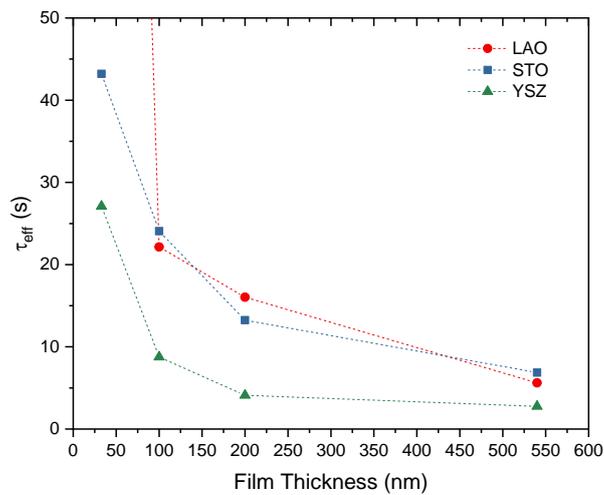

Figure 8: Thickness dependence of the exchange time constants of oxidation at 375 °C.

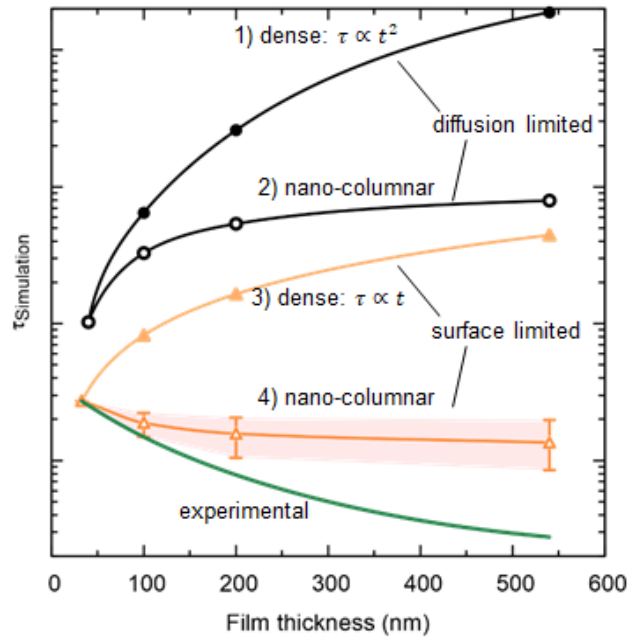

Figure 9: Expected thickness dependence of the saturation time, $\tau$ (as obtained from ECR measurements) for dense (open symbols) and nano-columnar (filled symbols) thin film structures for the surface (triangles) and diffusion limited (circles) regime, obtained from finite element simulations. Model description and parameters, used for the simulations, are given in the main text and Table 1 and correspond to experimentally observed values. Scenario 4), regarding the nano-columnar structured, surface limited case, strongly depends on the grain size and its evolution with thickness. The green line corresponds to the experimentally obtained thickness dependence for L2NO4/YSZ (data from Figure 8).

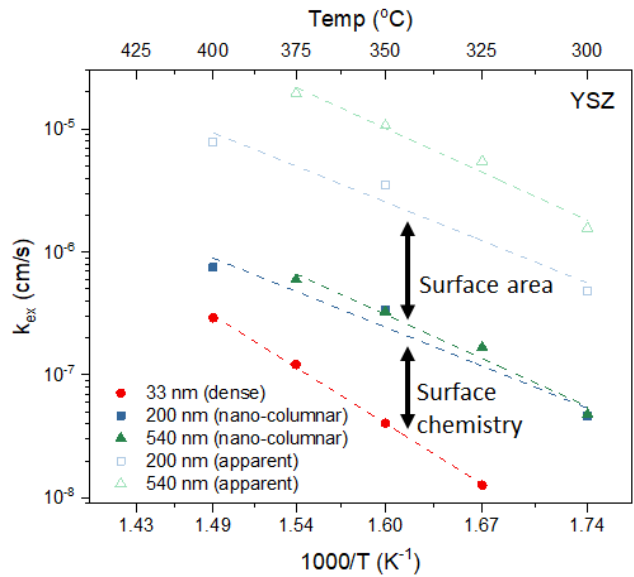

Figure 10: Effect of nano-columnar growth on the surface exchange coefficient, $k_{ex}$ of L2NO4/YSZ: tailoring the microstructure from dense (33 nm) to nano-columnar (200 and 540 nm) leads to a significant enhancement of $k_{ex}$, due to modifications of the surface chemistry. The influence of the enlarged surface area is depicted using the apparent exchange coefficient, equivalent to the surface exchange coefficient required for a dense layer to exchange oxygen as fast as a nano-columnar one.

# Supplementary Information

# Tailored nano-columnar La$_2$NiO$_4$ cathodes for improved electrode performance

Alexander Stangl *[a], Adeel Riaz [a], Laetitia Rapenne [a], José Manuel Caicedo [b], Carmen Jiménez [a], Michel Mermoux [c] and Mónica Burriel *[a]

[d.] Univ. Grenoble Alpes, CNRS, Grenoble INP, LMGP, 38000 Grenoble, France.
E-mail: alexander.stangl@grenoble-inp.fr and monica.burriel@grenoble-inp.fr
[e.] Catalan Institute of Nanoscience and Nanotechnology, ICN2, CSIC and The Barcelona Institute of Science and Technology (BIST), 08193 Bellaterra, Spain
[f.] Univ. Grenoble Alpes, Univ. Savoie Mont Blanc, CNRS, Grenoble INP, LEPMI, 38000, Grenoble, France

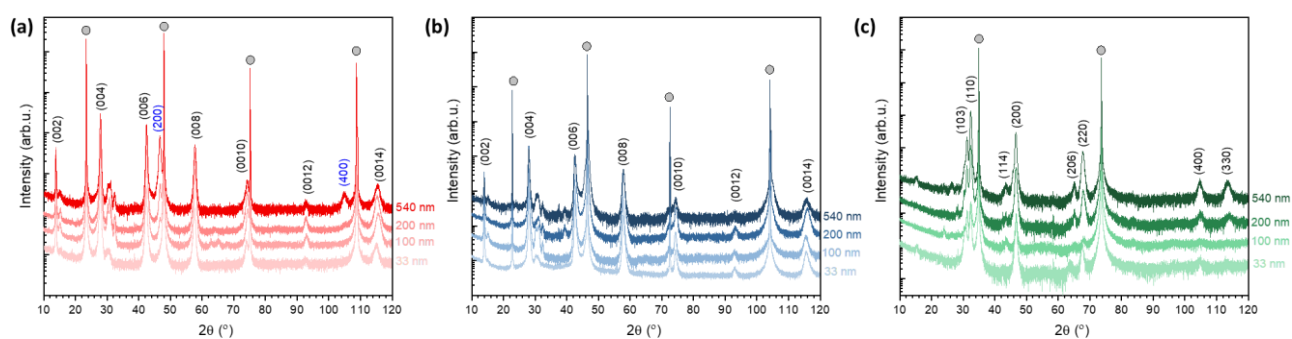

ESI-Figure 1: XRD pattern of L2NO4 thin films with different thickness deposited at 650 °C on a) LAO, b) STO and c) YSZ. Substrate peaks are marked by grey dots and the observed L2NO4 planes are labelled according to ICDD reference 04-015-2147.

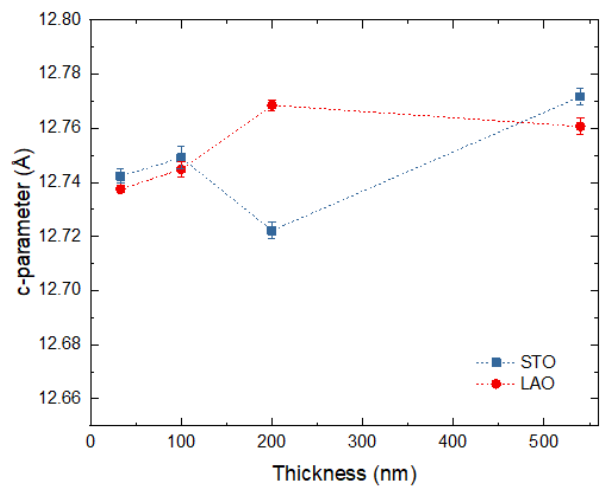

ESI-Figure 2: *c*-parameter as a function of thickness for "as-deposited" L2NO4 films on LAO and STO single crystal substrates.

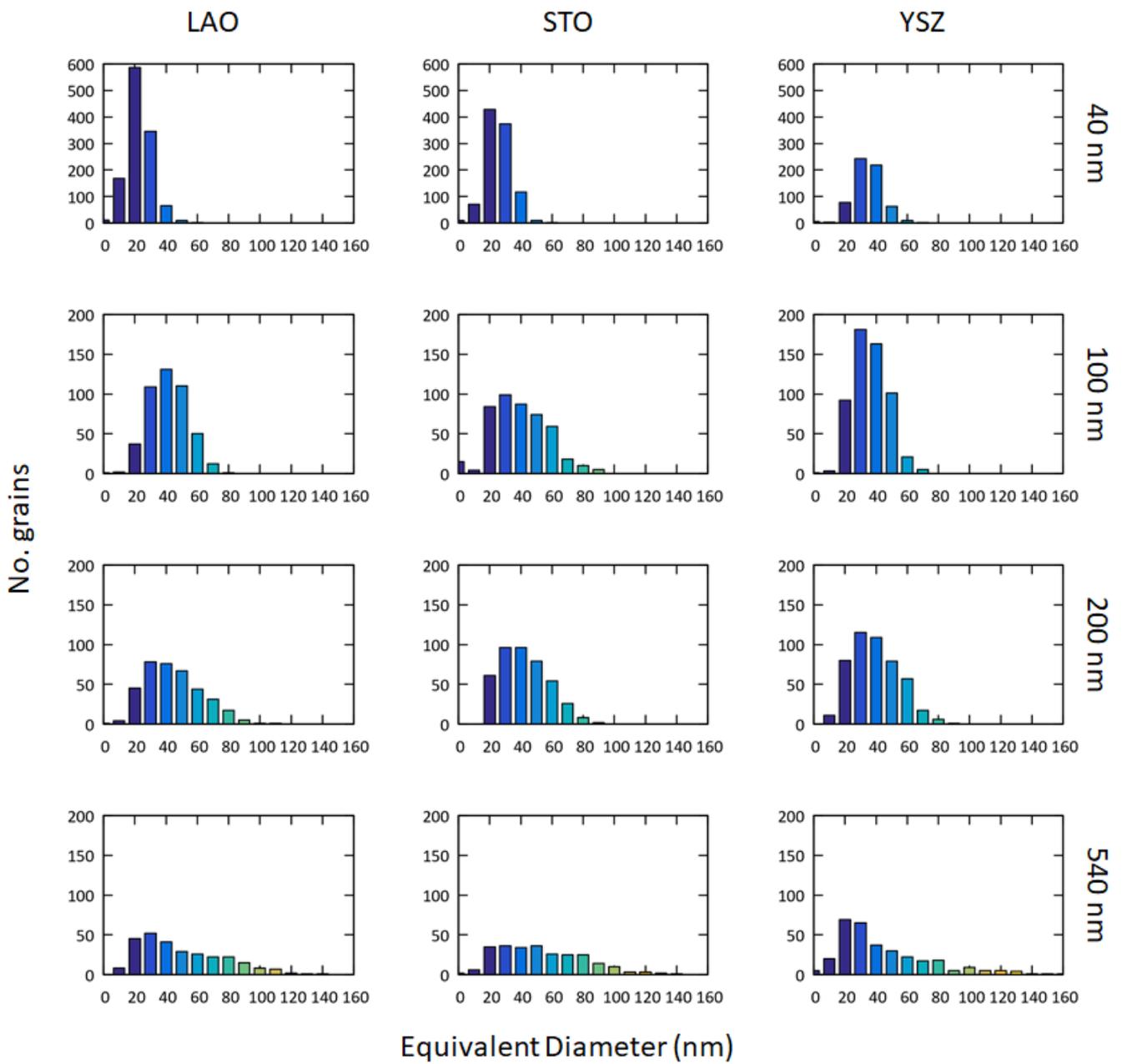

ESI-Figure 3: Grain size distribution, i.e. number of grains for each grain size, for all L2NO4 film thicknesses and substrates, analysed within an area of 1.13×0.76 μm.

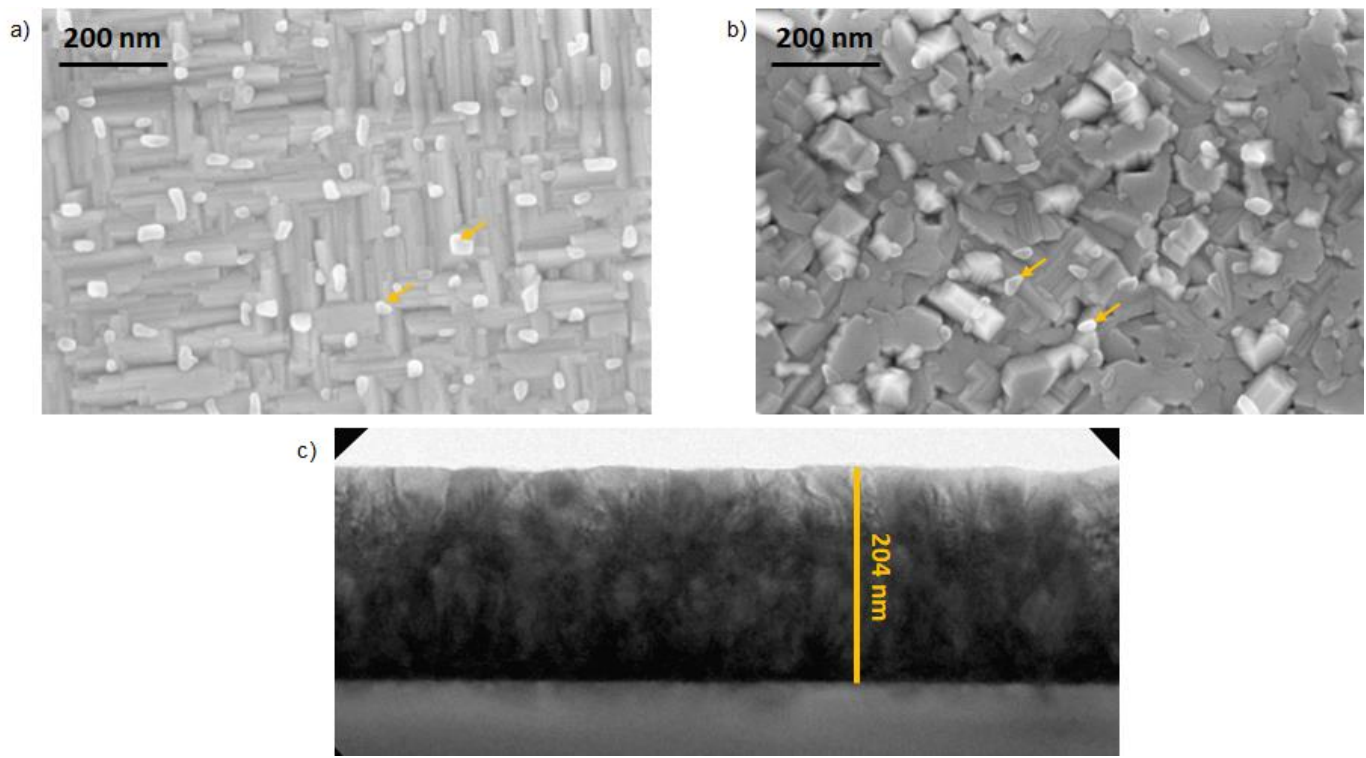

ESI-Figure 4: SEM analysis of 200 nm thick L2NO4 film deposited at 750 °C on a) LAO and b) YSZ substrate. Deposition at elevated temperature results in a dense structure with a closed, flat surface, as confirmed by TEM analysis, shown in c) for the L2NO4/LAO sample. Segregation of excess La on the surface is observed in the form of $La_2O_3$ particles (yellow arrows). The La/Ni ratio in the precursor solution was optimised for depositions at 650 °C. The solution-to-layer transfer ratio however depends on the deposition temperature. The La segregation can be avoided by reducing the La/Ni ratio in the precursor solution. However, no influence of excess La was found on the exchange activity.

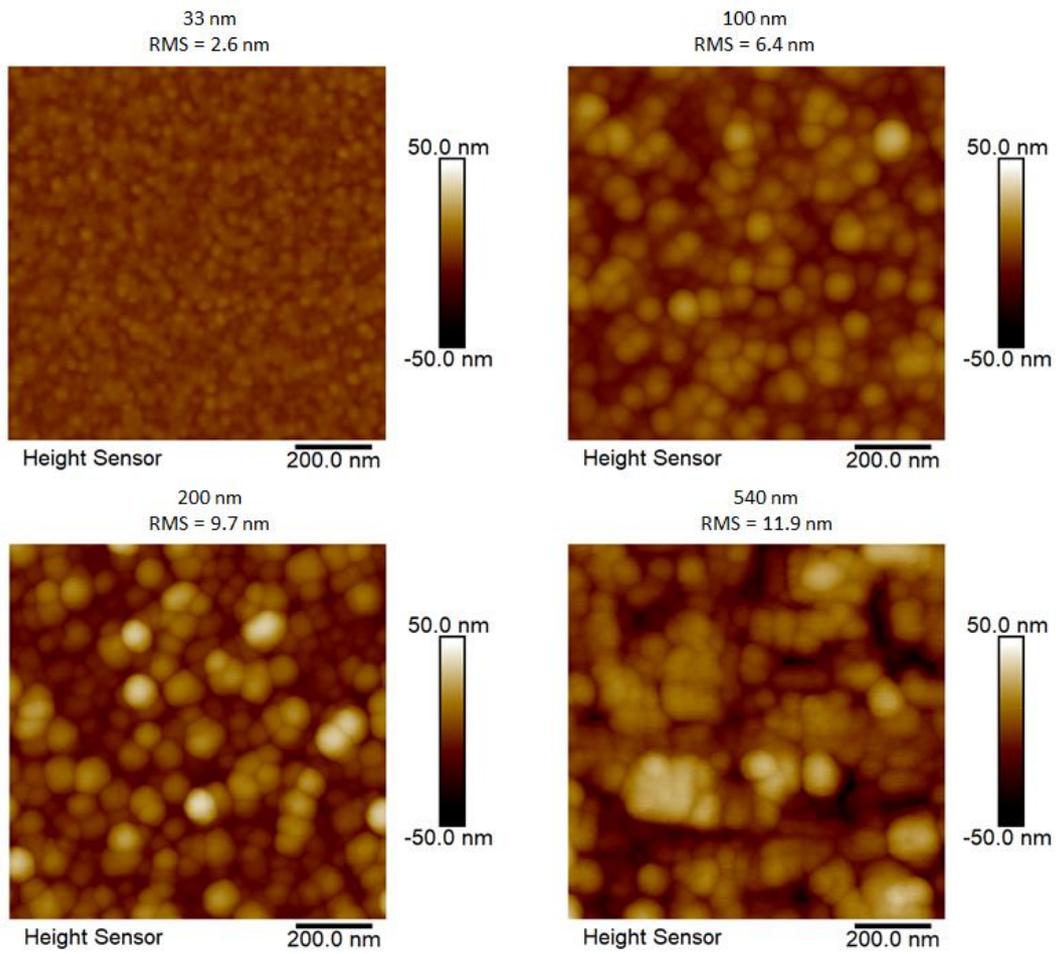

ESI-Figure 5: Atomic force microscopy of L2NO4/LAO samples, revealing increasing roughness (RMS) from 2.6 to 11.9 nm with increasing film thickness from 33 to 540 nm.

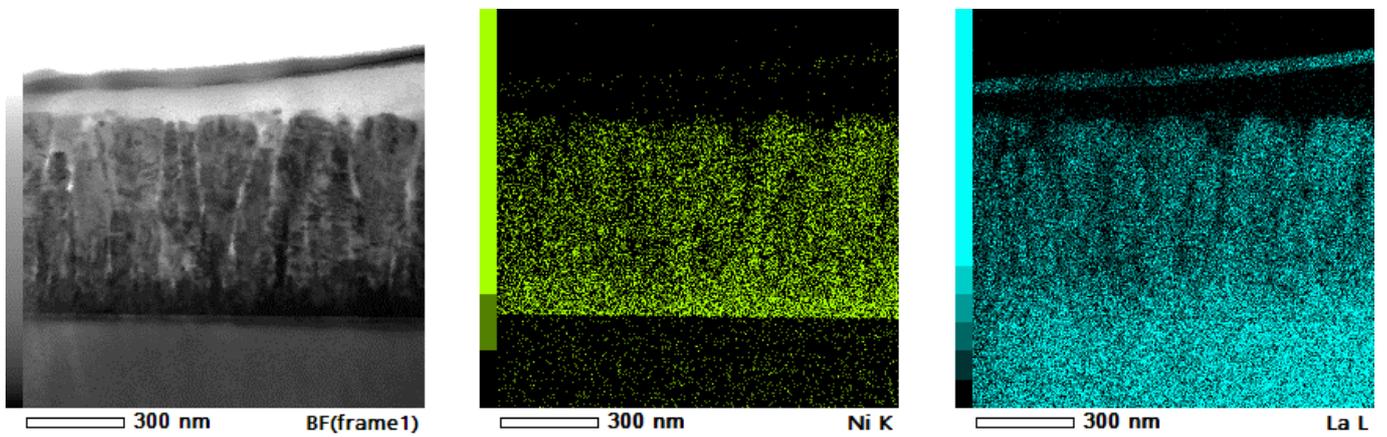

ESI-Figure 6: STEM EDX analysis of 540 nm thick L2NO4/LAO sample.

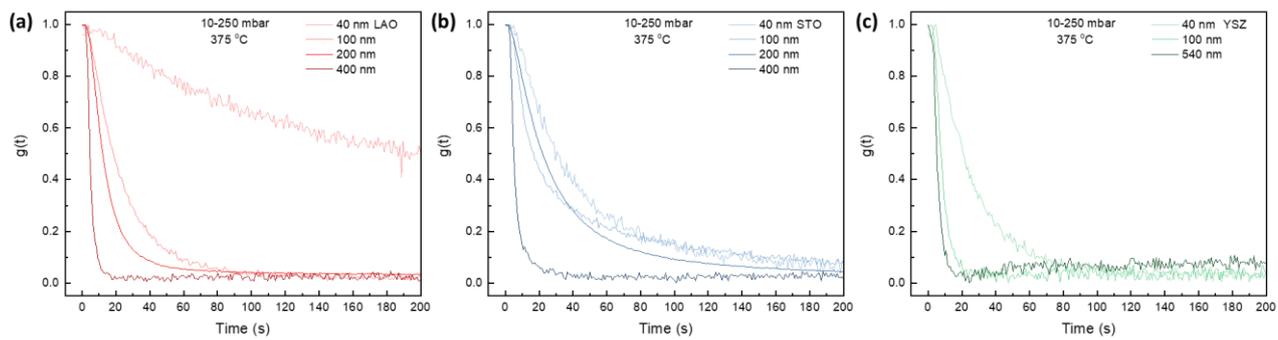

ESI-Figure 7: Normalised conductivity transients of L2NO4 thin films at 375°C after a change of $pO_2$ from 10-250 mbar. Films of different thickness deposited on a) LAO, b) STO and c) YSZ.